\documentclass{ws-procs975x65}

\usepackage{graphicx}
\usepackage{amsmath}
\usepackage{amsfonts}
\usepackage{amssymb}
\usepackage{amsbsy}

\newcommand{\Hil}{\mathcal{H}}

\def\p {\partial}

\newcommand{\rd}{{\rm d}}

\begin{document}

\title{A COMPUTABLE FRAMEWORK FOR LOOP QUANTUM GRAVITY}

\author{VIQAR HUSAIN$^*$ and TOMASZ PAW{\L}OWSKI$^\dagger$}

\address{$^*$Department of Mathematics and Statistics, University of New Brunswick,\\
  Fredericton, NB  E3B 2S9, Canada \\
  E-mail: vhusain@unb.ca\\
  $^\dagger$ Department of Mathematical Methods in Physics, Faculty of Science, University of Warsaw\\
  Warsaw, Poland\\
  E-mail: tomasz.pawlowski@fuw.edu.pl
}

\begin{abstract}
We present a  non-perturbative quantization of general relativity coupled to dust and other matter fields. The dust provides a natural time variable, leading to a physical  Hamiltonian with spatial diffeomorphism symmetry.  The methods of loop quantum gravity applied to this model lead to  a physical Hilbert space and Hamiltonian.  This provides a framework for  physical calculations  in the theory. 

\end{abstract}


\vskip 1.0 cm

\noindent {\bf Introduction} The goal of finding a background independent quantum theory of gravity with which one can begin to compute physics remains a central challenge of theoretical physics. Background independence  means that  the approach to quantization should not use a fixed metric on the physical manifold.  This situation is unlike that in standard quantum field theory or string theory where the Minkowski (or other) fixed metric, with its accompanying  symmetries, is used in a fundamental way 
in the quantization procedure. 

Loop quantum gravity (LQG) is one attempt  at a background independent approach. Its starting point is  classical general relativity possibly coupled to matter. Construction of a  canonical quantum theory  may proceed along one of two alternative paths: Dirac quantization, where all classical constraints are implemented as quantum constraints, or deparametrization using matter reference systems,\cite{bk} where gauge conditions are fixed classically to obtain a physical Hamiltonian. 
  
Both approaches continue to attract interest, but face significant issues. In the first, the problem of closure of the quantum constraint algebra is central; in the second, the reduced Hamiltonians obtained by full deparametrization using dust \cite{Giesel-Thiemann} or scalar fields are square roots, \cite{dgkl-qg} making it difficult to find operator realizations. 

In the approach we follow,\cite{HP-prl}  the first problem does not appear, and the second is solved.  This leads to a tractable system at the quantum level: a physical Hilbert space on which the Hamiltonian  operator can be  defined using LQG methods. The result is an in principle computable model for  quantum gravity in four dimensions.     
The approach makes only a time deparametrization, and makes use of the Hilbert space of diffeomorphism invariant states on which the physical Hamiltonian is defined.
  
\noindent {\bf Classical theory} The theory is given by the action \cite{HP-prl} 
\begin{equation}
  S = \frac{1}{4G} \int \rd^4x \sqrt{-g} R  - \frac{1}{2} \int  \rd^4x\  \sqrt{-g} M ( g^{ab}\partial_a T \partial_b T + 1) +\ S_{\rm SM} . \label{action}
  \end{equation}
The dynamical fields are  the metric $g_{ab}$, a dust field $T$,  and any other matter fields in the last term, $S_{\rm SM}$.  The second term is the dust action ($S_D$). It  
resembles the Brown-Kuchar  action \cite{bk}, but is different in that the dust is  irrotational.  The dust stress-energy tensor is
\begin{equation}
  T^{ab} = M U^a U^b + (M/2) g^{ab}\left( g_{cd} U^cU^d +1 \right),
  \label{Tab}
\end{equation}
where $U_a = \p_aT$; this is the usual form of dust field with rest mass $M$.  

The canonical decomposition of this theory gives the Hamiltonian and diffeomorphism constraints
\begin{eqnarray}\label{eq:Hg}
&& {\cal H}_G + {\cal H}_{SM} +  {\rm sgn} (M)  \left(p_T^2 + q^{ab}C_a^DC_b^D\right)^{1/2}= 0, \\
&&  C_{a}^G + C_{a}^{SM} + C_a^D=0,
\end{eqnarray} 
where $ {\cal H}_G , {\cal H}_{SM}, C_{a}^G ,C_{a}^{SM}$ are the gravitational and matter contributions to the respective constraints, $P_T$ is the momentum conjugate to $T$,
and $C_a^D =P_T\p_aT$.  
 
\noindent {\bf Deparametrization}  We note that (i) the requirement of the positivity of the mass in (\ref{Tab}) select $M>0$ and (ii) the gauge choice $T=t$ gives $C_a^D =0$.  {\it This leads to a remarkable simplicity: } the Hamiltonian constraint may be  solved strongly to give the physical Hamiltonian density 
\begin{equation}
{\cal H}_{phys}:=-p_T =  {\cal H}_G + {\cal H}_{SM}. 
\end{equation} 
 The spatial diffeomorphism constraint remains as 
\begin{equation}
 C_{a}^G + C_{a}^{SM}=0.
\end{equation}
 The last two equations  are the complete specification of the time deparametrized theory. It resembles gauge theory in that there is a physical hamiltonian and a set of first class constraints that form a Lie algebra.\footnote{This deparametrization was independently considered at the classical level by Kuchar and Torre.\cite{kuchar-torre}}
  
\noindent {\bf Quantization}  We set ${\cal H}_{SM}=0$ and describe the quantization of the system in the LQG framework. The canonical variables for gravity are 
the pair  $(A_a^i, E^a_i)$ where $A_a^i=\Gamma^i_a(E) + \gamma K_a^i$ is an su(2) connection, $E^{ai}$ is a vector density of weight one, and $K_a^i$ is the extrinsic curvature.  
The physical Hamiltonian density and constraints are  
\begin{eqnarray} \label{eq:Ham-class}
  {\cal H}_{phys} &=& \frac{\gamma^2}{2 \sqrt{{\rm det} E}}E^a_iE^b_j\left(\epsilon^{ij}_{\ \ k} F_{ab}^k + 2(1-\gamma^2)
  K_{[a}^iK_{b]}^j \right),\\
   {\cal G}_i &:=& \p_a E^a_i + \epsilon^k_{\ ij} A_a^j E_k^a \label{gauss}=0, \\
  C_a^G &:=&  E^b_i F_{ab}^i - A_a^i{\cal G}_i. \label{diffeo}=0.
 \end{eqnarray}

The phase space variables used for quantization are the holonomy
$h_\gamma(A) \equiv P\exp \int_\gamma A_a^i \tau^i \rd x^a$ and the flux
$K^i = \int_S E^{ai}\rd\sigma_a$, where the loop $\gamma$ and surface $S$ are  embedded in a spatial slice, and $\tau^i$ is a generator of the group. The Poisson bracket of these variables is called the holonomy-flux algebra.  

This algebra is quantized on the  Hilbert space {$\Hil_{\rm kin}$} spanned by the spin-networks states $|\Gamma; j_1\cdots j_m; I_1\cdots I_m\rangle$\cite{Thiemann-book}; the basis is labelled by three sets of quantum numbers: a graph $\Gamma$ embedded in the 3-manifold,  an assignment of spin labels $j$ on its  edges, and  intertwiners $I$ on its vertices (which sew together the spins entering a vertex).

We now note that  there is a construction of the Hilbert space of gauge and diffeomorphism invariant states, ${\cal H}_{diff}$,  obtained by solving the Gauss constraint (\ref{gauss}) by projection in ${\cal H}_{kin}$, and group averaging finite diffeomorphisms.  These two steps convert (classes of)  fixed  topology embedded graphs into  abstract graphs labelled  only by  spins  and  intertwiners;  this characterizes  the basis  $|j_1 \cdots j_N, I_1\cdots I_M\rangle$ of  ${\cal H}_{diff}$, which is {\it the physical Hilbert space of our theory}. The kinematical area and volume operators of LQG become physical observables in our theory.  

It remains to construct on ${\cal H}_{diff}$ the operator for the physical Hamiltonian $H_{phys} = \int d^3x {\cal H}_G $.  This is done by  using the basic  holonomy $\hat{h}_\gamma$ and volume $\hat{V}$  operators\cite{Thiemann-book}, from which composite operators  in  ${\cal H}_G$ are constructed. One  chooses a fixed graph $\alpha$\footnote{The choice includes (but is not restricted to) the tetrahedral or cubic lattice.}. The operator is a sum over the graph vertices  $\hat{H}^G=\sum_{v\in V(\alpha)}\hat{\mathcal{H}}^G_v$, where $\hat{\mathcal{H}}^G_v$ is  composed of $(i)$ the volume operator $\hat{V}(v)$, $(ii)$ the combination $\hat{h}_e[\hat{h}_e^{-1},\hat{V}]$ with holonomies  $\hat{h}_e$ along the adjacent edge, and $(iii)$ the holonomies $\hat{h}_{\square}(v)$ along the minimal closed loops in $\alpha$. The operators $(i)$ and $(ii)$ are diagonal in the basis, and their properties  are well understood; the operator in $(iii)$ changes the spin labels on the edges composing the loop. This construction is similar to that of Thiemann, except that the operator is now a physical Hamiltonian, and not a constraint.

\noindent {\bf Summary} We have given a background independent quantization of the theory (\ref{action}), with a specification of the physical Hilbert space and Hamiltonian operator.  It provides one completion of the LQG program with dust matter. It is now possible to set up calculations of actual physical effects on  fixed abstract graphs.

\bibliographystyle{ws-procs975x65}
\bibliography{MG13-qg}

\end{document}